\documentclass[conference]{IEEEtran}
\IEEEoverridecommandlockouts
\usepackage{cite}
\usepackage{amsmath,amssymb,amsfonts}
\usepackage{algorithmic}
\usepackage{graphicx}
\usepackage{tabularx}
\usepackage{textcomp}
\usepackage{xcolor}
\usepackage{stfloats}
\usepackage{multirow}
\usepackage{epstopdf}

\def\BibTeX{{\rm B\kern-.05em{\sc i\kern-.025em b}\kern-.08em
    T\kern-.1667em\lower.7ex\hbox{E}\kern-.125emX}}
\begin{document}

\title{A General 3D Non-Stationary Massive MIMO GBSM for 6G Communication Systems}

\author{Yi Zheng\textsuperscript{1,2}, Long Yu\textsuperscript{1,2}, Runruo Yang\textsuperscript{1,2}, and Cheng-Xiang Wang\textsuperscript{1,2*}
\\
\textsuperscript{1}National Mobile Communications Research Laboratory, School of Information of Science and Engineering, 
\\Southeast University, Nanjing 210096, China.\\

%\textsuperscript{3}State Key Laboratory of Millimeter Waves, School of Information of Science and Engineering, 
%\\Southeast University, Nanjing 210096, China.\\
\textsuperscript{2}Purple Mountain Laboratories, Nanjing 211111, China.\\
\textsuperscript{*}Corresponding Author: Cheng-Xiang Wang
\\
Email: \{zheng\_yi, yulong, yangrr, chxwang\}@seu.edu.cn
}

\maketitle

\begin{abstract}
A general three-dimensional (3D) non-stationary massive multiple-input multiple-output (MIMO) geometry-based stochastic model (GBSM) for the sixth generation (6G) communication systems is proposed in the paper. The novelty of the model is that the model is designed to cover a variety of channel characteristics, including space-time-frequency (STF) non-stationarity, spherical wavefront, spatial consistency, channel hardening, etc. Firstly, the introduction of the twin-cluster channel model is given in detail. Secondly, the key statistical properties such as space-time-frequency correlation function (STFCF), space cross-correlation function (CCF), temporal autocorrelation function (ACF), frequency correlation function (FCF), and performance indicators, e.g., singular value spread (SVS), and channel capacity are derived. Finally, the simulation results are given and consistent with some measurements in relevant literatures, which validate that the proposed channel model has a certain value as a reference to model massive MIMO channel characteristics.
\end{abstract}


	\section{Introduction}
	Compared with the fifth generation (5G) communication systems, the 6G communication systems have attracted more and more attention because of almost a thousand times transmission rate and capacity\cite{RE1}, \cite{RE23}. Massive MIMO technology is an efficient way to increase capacity and spectral efficiency for 6G communication systems, which refers to that the base station (BS) is equipped with a large number of antennas up to one hundred or even thousands of antennas. Besides, with more and more antennas exploited at BS side, the channel among different users become approximatively orthogonal, which is called channel hardening phenomenon (or favorable propagation conditions)\cite{RE2}, \cite{RE4}. Therefore, the interference among the users can be removed, which makes the 6G communication systems inherently robust.
	
	There are mainly three new characteristics for massive MIMO channel: spherical wavefront, spatial non-stationarity, and channel hardening phenomenon. Spherical wavefront refers to that the distance between the transmitter (Tx) and receiver (Rx) or cluster is less than the Rayleigh distance $2L^2/\lambda$, where $L$ represents the antenna array size, $\lambda$ denotes the wavelength. Channel measurements showed that the angle of departure (AoD) along the antenna array gradually shifts\cite{RE5} and line-of-sight (LOS) path azimuth angle shifts along the array\cite{RE6}, which demonstrated the spherical wavefront. Birth-death process along the array brings spatial non-stationarity. The clusters appear and disappear along the array randomly, which was verified by the fact that received power of LOS path varies along the array \cite{RE6}. In \cite{RE7}, theoretical analysis showed that the correlation matrix at user side becomes a diagonal matrix under favorable propagation conditions and channel hardening phenomenon is obvious. The above characteristics bring new requirements for massive MIMO channel modeling.
	
	%	Channel modeling is an efficient way to evaluate link-level, system-level, and network-level performance. Therefore, massive MIMO channel modeling is indispensable. 
	A general massive MIMO channel model for 6G communication systems should be at least suitable for millimeter wave communications, vehicle-to-vehicle (V2V) communications, 3D communication environments, and high-speed train (HST) communications.
	%	 In general, massive MIMO stochastic channel models can be classified into two categories: GBSM and correlation-based stochastic model (CBSM)\cite{RE8}. 
	%	 CBSMs contains Kronecker-based stochastic model (KBSM), virtual channel representation (VCR), and Weichselberger channel model. Because GBSM is more accurate than CBSM, it has been widely used in engineering field.
	In \cite{RE9}, a two-dimensional (2D) parabolic wavefront model was proposed, and a 3D parabolic wavefront model was further developed \cite{RE10}. However, the model only considered the movements of the Rx and clusters, so it was not applicable to V2V scenario. Similarly, a twin-cluster channel model in \cite{RE11} did not consider the movement of the Tx. A general 3D non-stationary 5G channel model and a general 3D non-stationary channel model for 5G and beyond were given in \cite{RE12} and \cite{RE13}, respectively. The two models only considered the cluster evolution in space domain and time domain, which was not suitable for millimeter wave communication systems needing to consider cluster evolution in frequency domain. References \cite{RE14} and \cite{RE15} demonstrated a multi-ring channel model and a multi-confocal ellipse channel model, respectively. Both of them were 2D channel models without considering elevation characteristic. Reference \cite{RE16} proposed a 3D ellipsoid model, which did not consider the movement of the Tx and cluster. Reference \cite{RE17} proposed a millimeter wave massive MIMO channel for HST communications without considering spatial consistency and V2V communications. The above mentioned channel models do not consider all the requirements for 6G massive MIMO channel modeling.

	To the best of our knowledge, the general 3D massive MIMO GBSM considering STF cluster evolution, spherical wavefront, spatial consistency, and channel hardening for 6G communication systems is still missing in the literature. This paper presents a general 3D massive MIMO GBSM based on the model in \cite{RE13} so as to fill the above research gaps. The contributions of the paper are summarized as follows. Firstly, the proposed channel model is suitable for millimeter wave communications, V2V communications, 3D communication environments, and HST communications by adjusting channel parameters. Secondly, the cluster evolution is further extended from space and time domain in \cite{RE13} to STF domain. Thirdly, the large scale parameters (LSPs) and small scale parameters (SSPs) are generated according to the positions of Tx and Rx using the sum-of-sinusoids (SoS) method, which makes the model inherently spatially consistent. Finally, the STF non-stationarity, spatial consistency, and channel hardening characteristics are verified by simulation results. 
	%The uplink sum-rates increase with the number of antennas proved that the massive MIMO channel model can improve the capacity significantly and can be applied to 6G communication systems.

	The remaining paper is structured as follows.
	Section~\uppercase\expandafter{\romannumeral2} describes the proposed general massive MIMO GBSM in detail. In Section~\uppercase\expandafter{\romannumeral3}, statistical properties 
	%, e.g., STFCF, space CCF, temporal ACF, FCF, 
	and performance indicators of the presented model
	%such as SVS, and channel capacity 
	are derived. Simulation results and analysis are given in Section~\uppercase\expandafter{\romannumeral4}. Finally, conclusions are drawn in Section~\uppercase\expandafter{\romannumeral5}.

	\section{A general Massive MIMO GBSM}

\begin{figure}[b]
		\centerline{\includegraphics[width=0.48\textwidth]{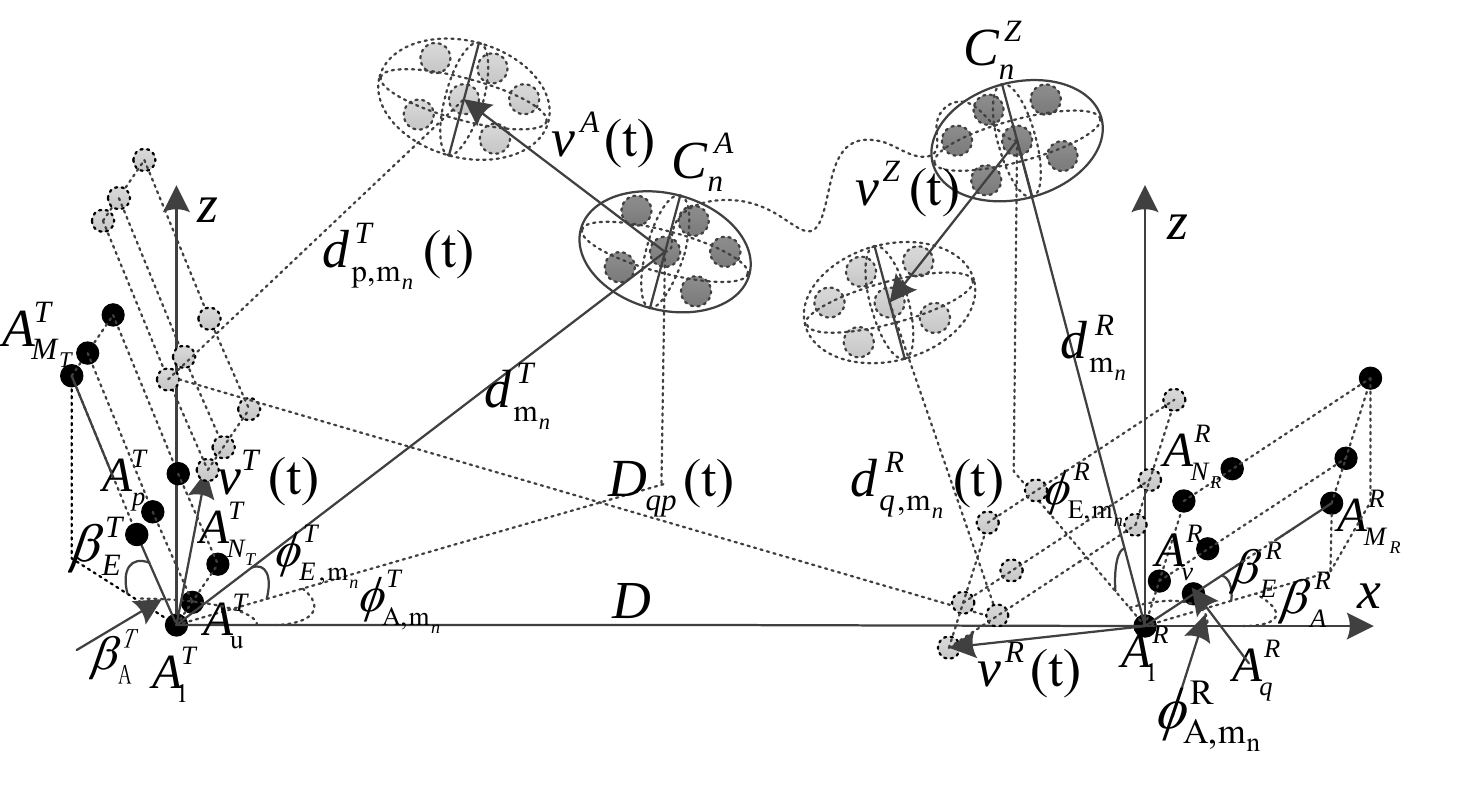}}
		\caption{A general 3D massive MIMO GBSM for 6G communication systems.}
		\label{fig_1}
	\end{figure}

	As illustrated in Fig. \ref{fig_1}, large uniform rectangular arrays (URAs) are adopted at the BS and mobile station (MS) sides in this model. Suppose that the BS is Tx and the MS is Rx. The URA at BS (MS) side is formed by uniform linear arrays (ULAs) in two dimensions. There are $M_T$ ($M_R$) antenna elements symboled as $A_p^T$ ($p=1,2,\cdots, M_T$) ( $A_q^R$ ($q=1,2,\cdots, M_R$)) and spaced at a distance $\delta_T$ ($\delta_R$) in one dimension. In another dimension, there are $N_T$ ($N_R$) antenna elements symboled as $A_u^T$ ($u=1,2,\cdots, N_T$) ($A_v^R$ ($v=1,2,\cdots, N_R$)) spaced at a distance $\delta_T$ ($\delta_R$). 
	%The URA at MS side is similar to the URA at BS side. There are $M_R$ antenna elements symboled as $A_q^R$ ($q=1,2,\cdots, M_R$) and spaced at a distance $\delta_R$ in one dimension. In another dimension, there are $N_R$ antenna elements symboled as $A_v^R$ ($v=1,2,\cdots, N_R$) and spaced at a distance $\delta_R$. 
	In the $M_T$ ($M_R$) antenna elements dimension, the angle of elevation is $\beta_E^T$ ($\beta_E^R$) and the angle of azimuth is $\beta_A^T$ ($\beta_A^R$). %In the $N_T$ ($N_R$) antenna elements dimension, the angle of elevation is $\zeta_E^T$ ($\zeta_E^R$) and the angle of azimuth is $\zeta_A^T$ ($\zeta_A^R$). 
	In order to calculate conveniently, we consider the $M_T$ ($M_R$) antenna elements as a ULA. All the ULAs in the $N_T$ ($N_R$) antenna elements dimension  can be added to form the whole URA. Multi-bounce propagation is simplified as twin-cluster propagation. The path between the first bounce cluster $C_n^A$ and the last bounce cluster $C_n^Z$ is abstracted by a virtual link. The total number of paths from $A_p^T$ to $A_q^R$ at time $t$ is $N_{qp}(t)$. The number of scatterers in the $n$th path is $M_n(t)$. The Tx, Rx, and clusters can move with arbitrary velocities and trajectories. Furthermore, all the parameters are time-variant. For clarity, the remaining definitions of the parameters are shown in Table~ \uppercase\expandafter{\romannumeral1}.

	\renewcommand{\arraystretch}{1.3}
	\begin{table*}[tb]
		\centering
		\caption{Definition of Key Channel Model Parameters.}
		\begin{tabular}{|c|c|}			
			\hline
			\textbf{Parameters}&\textbf{Definition}\\
			\hline
			$f_c$&Carrier frequency
			\\\hline	
			$D$&Distance from $A_1^T$ to $A_1^R$ at initial time
			\\\hline
			$d_{m_n}^{T(R)}$&Distance from $A_1^{T(R)}$ to the $m$th scatterer in $C_n^{A(Z)}$ at initial time
			\\\hline
			$d_{p(q),m_n}^{T(R)}(t)$&Distance from $A_{p(q)}^{T(R)}$ to the $m$th sactterer in $C_n^{A(Z)}$ at time $t$
			\\\hline
			$d_{qp,m_n}(t)$&Distance from $A_p^T$ through the $m$th scatterer in $C_n^A$ and the $m$th scatterer in $C_n^Z$ to $A_q^R$ at time $t$
			\\\hline	
			$\tau_{qp,m_n}(t)$&Delay from $A_p^T$ through the $m$th scatterer in $C_n^A$ and the $m$th scatterer in $C_n^Z$ to $A_q^R$ at time $t$
			\\\hline
			$v^T(t)$, $v^R(t)$, $v^{A_n}(t)$, $v^{Z_n}(t)$&Speeds of the Tx, Rx, cluster $C_n^A$, and cluster $C_n^Z$ at time $t$
			\\\hline
			$\alpha_{A}^T(t)$, $\alpha_{A}^R(t)$, $\alpha_{A}^{A_n}(t)$,$\alpha_{A}^{Z_n}(t)$&Azimuth angles of movements of the Tx, Rx, cluster $C_n^A$, and cluster $C_n^Z$ at time $t$
			\\\hline
			$\alpha_{E}^T(t)$, $\alpha_{E}^R(t)$, $\alpha_{E}^{A_n}(t)$,$\alpha_{E}^{Z_n}(t)$&Elevation angles of movements of the Tx, Rx, cluster $C_n^A$, and cluster $C_n^Z$ at time $t$
			\\\hline
			$\phi_{A,{LOS}}^T$, $\phi_{E,{LOS}}^T$&Azimuth angle of departure (AAoD) and elevation angle of departure (EAoD) from $A_1^T$ to $A_1^R$ at initial time
			\\\hline			
			$\phi_{A,{LOS}}^R$, $\phi_{E,{LOS}}^R$&Azimuth angle of arrival (AAoA) and elevation angle of arrival (EAoA) from $A_1^R$ to $A_1^T$ at initial time
			\\\hline	
			$\phi_{A,m_n}^{T(R)}$, $\phi_{E,m_n}^{T(R)}$&AAoD (AAoA) and EAoD (EAoA) from $A_1^{T(R)}$ to the $m$th scatterer in $C_n^{A(Z)}$ at initial time
			%	\\\hline
			%	$\phi_{A,m_n}^R$, $\phi_{E,m_n}^R$&AAoA and EAoA from $A_1^R$ to the $m$th scatterer in $C_n^Z$ at initial time
			\\\hline	
			
			$P_{qp,m_n}(t)$&Power of the ray from $A_p^T$ through the $m$th scatterer in $C_n^A$ and the $m$th scatterer in $C_n^Z$ to $A_q^R$ at time $t$
			\\\hline		
		\end{tabular}
		
		\label{tab1}
	\end{table*}
	
	%	
	%\begin{figure*}[tb]
	%	\centerline{\includegraphics[width=\textwidth]{Pa.png}}
	%%	\caption{A general Massive MIMO GBSM for 6G communication systems.}
	%%	\label{fig_1}
	%\end{figure*}	
	
	\subsection{Channel Impulse Response (CIR)}
	The complete channel matrix is comprised of large scale fading (LSF) part and small scale fading (SSF) part. The LSF consists of path loss ($PL$), shadowing ($SH$), blockage loss ($BL$), and gas absorption loss ($AL$). The theoretical channel matrix is presented as
	\begin{equation}\label{H}
		\textbf{H}=\left[PL\cdot
		SH\cdot
		BL\cdot
		AL
		\right]^{1/2}\textbf{H}_s
	\end{equation} where $\textbf{H}_s$ is the SSF matrix and can be further represented as
	\begin{equation}
		%	\label{H_s}
		\textbf{H}_s=\left[h_{qp}(t,\tau)\right]_{M_RN_R\times
			M_TN_T}
	\end{equation}
	\noindent where $h_{qp}(t,\tau)$ can be acquired by the summation of LOS component and non-line-of-sight (NLOS) components.
	\begin{equation}\label{h}
		h_{qp}(t,\tau)=\sqrt{\frac{K_{RF}(t)}
			{K_{RF}(t)+1}}h_{qp}^{L}(t,\tau)+\sqrt{\frac{1}
			{K_{RF}(t)+1}}h_{qp}^{N}(t,\tau)
	\end{equation} where $K_{RF}(t)$ is Rician factor, $h_{qp}^{L}(t,\tau)$ is LOS component and $h_{qp}^{N}(t,\tau)$ NLOS components. LOS component can be represented as
	\begin{equation}\label{h_L}
		\begin{split}
			&h_{qp}^L(t,\tau)
			=\left[
			\begin{matrix}
				F_{q,V_p}(\phi_{E,{LOS}}^R(t), \phi_{A,{LOS}}^R(t))\\
				F_{q,H_p}(\phi_{E,{LOS}}^R(t), \phi_{A,{LOS}}^R(t))
			\end{matrix}
			\right]^T\\
			&\cdot \left[
			\begin{matrix}
				e^{j\theta_{LOS}^{V_pV_p}}&0\\
				0&-e^{j\theta_{LOS}^{H_pH_p}}
			\end{matrix}
			\right]
			\left[
			\begin{matrix}
				F_{p,V_p}(\phi_{E,{LOS}}^T(t), \phi_{A,{LOS}}^T(t))\\
				F_{p,H_p}(\phi_{E,{LOS}}^T(t), \phi_{A,{LOS}}^T(t))
			\end{matrix}
			\right]\\
			& \cdot e^{j2\pi
				f_c\tau_{qp}^{LOS}(t)}\delta(\tau-\tau_{qp}^{LOS}(t)).
		\end{split}
	\end{equation}
	NLOS components can be represented as
	\begin{equation}\label{h_N}
		\begin{split}
			&h_{qp}^N(t,\tau)
			=\sum_{n=1}^{N_{qp}(t)}\sum_{m=1}^{M_{n}(t)}\left[
			\begin{matrix}
				F_{q,V_p}(\phi_{E,m_n}^R(t), \phi_{A,m_n}^R(t))\\
				F_{q,H_p}(\phi_{E,m_n}^R(t), \phi_{A,m_n}^R(t))
			\end{matrix}
			\right]^T\\
			&\cdot \left[
			\begin{matrix}
				e^{j\theta_{m_n}^{V_pV_p}}& \sqrt{\kappa_{m_n}^{-1}(t)}e^{j\theta_{m_n}^{V_pH_p}}\\
				\sqrt{\kappa_{m_n}^{-1}(t)}e^{j\theta_{m_n}^{H_pV_p}}&e^{j\theta_{m_n}^{H_pH_p}}
			\end{matrix}
			\right]\\
			&\cdot \left[
			\begin{matrix}
				F_{p,V_p}(\phi_{E,m_n}^T(t), \phi_{A,m_n}^T(t))\\
				F_{p,H_p}(\phi_{E,m_n}^T(t), \phi_{A,m_n}^T(t))
			\end{matrix}
			\right]
			\sqrt{P_{qp,m_n}(t)}e^{j2\pi
				f_c\tau_{qp,m_n}(t)}\\
			&\cdot \delta(\tau-\tau_{qp,m_n}(t))
		\end{split}
	\end{equation} where $(\cdot)^T$ denotes the transpose operation. $F_{p(q),V_p}^{T(R)}(\cdot)$ and $F_{p(q),H_p}^{T(R)}(\cdot)$ represent the vertical polarization and horizontal polarization at Tx (Rx) side, respectively. $\kappa_{m_n}$ denotes the cross polarization ratio.
	\subsection{Channel Transfer Function (CTF)}
	Take the Fourier transform of the CIR, we will get the CTF as
	\begin{align}\label{}
		H_{qp}(t,f)=&\sqrt{\frac{K_{RF}(t)}{K_{RF}(t)+1}}H_{qp}^L(t,f)\notag\\
		&+\sqrt{\frac{1}{K_{RF}(t)+1}}H_{qp}^N(t,f).
	\end{align} 
	where $H_{qp}^L(t,f)$ is LOS component and $H_{qp}^N(t,f)$ NLOS components.
	%\begin{equation}\label{H_L}
	%\begin{split}
	% &H_{qp}^L(t,f)
	% =\left[
	% \begin{matrix}
	% F_{q,V_p}(\phi_{E,{LOS}}^R(t), \phi_{A,{LOS}}^R(t))\\
	% F_{q,H_p}(\phi_{E,{LOS}}^R(t), \phi_{A,{LOS}}^R(t))
	% \end{matrix}
	% \right]^T\\
	%&\left[
	%\begin{matrix}
	%e^{j\theta_L^{V_pV_p}}&0\\
	% 0&-e^{j\theta_L^{H_pH_p}}
	%\end{matrix}
	% \right]
	%  \left[
	%  \begin{matrix}
	%  F_{p,V_p}(\phi_{E,L}^T(t), \phi_{A,{LOS}}^T(t))\\
	% F_{p,H_p}(\phi_{E,L}^T(t), \phi_{A,{LOS}}^T(t))
	%  \end{matrix}
	% \right]\\
	% &e^{j2\pi
	% 	(f_c-f)\tau_{qp}^L(t)}
	% \end{split}
	%  \end{equation} 
	%and $H_{qp}^N(t,f)$ is obtained as
	%\begin{equation}\label{H_N}
	%\begin{split}
	%&H_{qp}^N(t,f)
	%=\sum_{n=1}^{N_{qp}(t)}\sum_{m=1}^{M_{n}(t)}\left[
	%\begin{matrix}
	% F_{q,V_p}(\phi_{E,m_n}^R(t), \phi_{A,m_n}^R(t))\\
	% F_{q,H_p}(\phi_{E,m_n}^R(t), \phi_{A,m_n}^R(t))
	% \end{matrix}
	%\right]^T\\
	% &\left[
	% \begin{matrix}
	% e^{j\theta_{m_n}^{V_pV_p}}& \sqrt{\kappa_{m_n}^{-1}(t)}e^{j\theta_{m_n}^{V_pH_p}}\\
	% \sqrt{\kappa_{m_n}^{-1}(t)}e^{j\theta_{m_n}^{H_pV_p}}&e^{j\theta_{m_n}^{H_pH_p}}
	% \end{matrix}
	% \right]\\
	%  &\left[
	% \begin{matrix}
	% F_{p,V_p}(\phi_{E,m_n}^T(t), \phi_{A,m_n}^T(t))\\
	% F_{p,H_p}(\phi_{E,m_n}^T(t), \phi_{A,m_n}^T(t))
	% \end{matrix}
	%\right]
	%\sqrt{P_{qp,m_n}(t)}e^{j2\pi
	% 	(f_c-f)\tau_{qp,m_n}(t)}.
	% \end{split}
	% \end{equation}
	\subsection{STF Cluster Evolution}
	The proposed channel has the characteristic of STF non-stationarity. Clusters may appear and disappear in STF domain. The space-time evolution is modeled jointly. For initial moment $t_i$ and antenna element $A_p^T$ ($A_q^R$), the cluster is represented as $C_p^T(t_i)$ ($C_q^R(t_i)$). At the next moment $t_i+\triangle t$, the cluster evolves into $C_{p+1}^T(t_i+\triangle t)$ ($C_{q+1}^R(t_i+\triangle t)$). The space-time evolution process can be modeled as    
	\begin{equation}\label{E_T}
		%\begin{split}
		\begin{matrix}
			\,\,\,
			C_p^T(t_i)\xrightarrow{E}
			C_{p+1}^T(t_i+\triangle t)&(p=1,2,\cdots, M_T-1)
		\end{matrix}
		%\end{split}
	\end{equation}    
	\begin{equation}\label{E_R}
		%\begin{split}
		\begin{matrix}
			\,\,\,
			C_q^R(t_i)\xrightarrow{E}
			C_{q+1}^R(t_i+\triangle t)&(q=1,2,\cdots, M_R-1).
		\end{matrix}
		%\end{split}
	\end{equation}  
	
	By defining $\lambda_G$ and $\lambda_R$ as the generation rate and recombination rate of the cluster, the survival probabilities of the clusters at Tx and Rx sides can be represented as
	\begin{equation}\label{E_T}
		\begin{split}
			\begin{matrix}
				\indent P_{\text{survival}}^T(\triangle t,\delta_p)=e^{-\lambda_R\left[(\epsilon_1^T)^2+(\epsilon_2^T)^2+2\epsilon_1^T\epsilon_2^T
					\text{cos}(\alpha_A^T-\beta_A^T)\right]^{1/2}}
			\end{matrix}
		\end{split}
	\end{equation}
	\begin{equation}\label{E_R}
		\begin{split}
			\begin{matrix}
				\indent P_{\text{survival}}^R(\triangle t,\delta_q)=e^{-\lambda_R\left[(\epsilon_1^R)^2+(\epsilon_2^R)^2+2\epsilon_1^R\epsilon_2^R
					\text{cos}(\alpha_A^R-\beta_A^R)\right]^{1/2}}
			\end{matrix}
		\end{split}
	\end{equation} where $\epsilon_1^T=\frac{\delta_p
	cos\beta_E^T}{D_c^A}, \delta_p=(p-1)\delta_T, $ ($\epsilon_1^R=\frac{\delta_q
	cos\beta_E^R}{D_c^A}, \delta_q=(q-1)\delta_R$), and $\epsilon_2^T=\frac{v_T\triangle t}{D_c^S}$ ($\epsilon_2^R=\frac{v_R\triangle t}{D_c^S}$) represent the distance differences caused by array evolution and time evolution, respectively. $D_c^A$ and $D_c^S$ are scenario-dependent coefficients in space domain and time domain, respectively. Combined with frequency evolution, the total survival probability is represented as\begin{equation}\label{E_To}
	\begin{split}
		%\begin{matrix}
		P_{\text{survival}}(\triangle t,\delta_p,\delta_q,\triangle f)&=P_{\text{survival}}^T(\triangle t,\delta_p)\cdot P_{\text{survival}}^R(\triangle t,\delta_q)\\
		&\cdot P_{\text{survival}}(\triangle f)
		%\end{matrix}
	\end{split}
\end{equation} 
where $P_{\text{survival}}(\triangle f)$ can be further represented as\cite{RE17}
\begin{equation}\label{E_F}
	%\begin{split}
	%\begin{matrix}
	P_{\text{survival}}(\triangle f)=e^{-\lambda_R\frac{F(\triangle f)}{D_c^f}}    
	%\end{matrix}
	%\end{split}
\end{equation} where $F(\triangle f)$ and $D_c^f$ are determined by channel measurements. Furthermore, the number of the clusters which are newly generated by STF evolution can be represented as
\begin{equation}\label{E_F}
	E\left[N_{\text{new}}\right]=\frac{\lambda_G}{\lambda_R}(1-P_{\text{survival}}(\triangle t,\delta_p,\delta_q,\triangle f)).
\end{equation}	

\section{Statistical Properties and Performance Indicator}
In this section, statistical properties and performance indicators of the proposed 3D massive MIMO GBSM are derived.
\subsection{The STFCF}
According to (6), we can get $H_{qp}(t,f)$ and $H_{q'p'}^*(t+\triangle t,f+\triangle f)$, then the STFCF can be defined as (16) at the bottom of the page, where $E\left[\cdot\right]$ defines expectation operation, and $(\cdot)^*$ defines the conjugation operation. $R_{qp,q'p'}^{L}(t,f;\triangle t,\triangle f,\delta_T,\delta_R)$ and $R_{qp,q'p'}^{N}(t,f;\triangle t,\triangle f,\delta_T,\delta_R)$ represent the STFCF of LOS component and NLOS components, respectively.
%	\begin{equation}
%	\begin{aligned}
%	&R_{qp,p'q'}(t,f;\triangle t,\triangle f,\delta_T,\delta_R)=\\
%&E\left[H_{qp}(t,f)H_{q'p'}^*(t+\triangle t,f+\triangle f)\right]=\\
%	&\sqrt{\frac{K_{RF}(t)}
%		{K_{RF}(t)+1}\cdot{\frac{K_{RF}(t+\triangle t)}
%			{K_{RF}(t+\triangle t)+1}}}R_{qp,p'q'}^{L}(t,f;\triangle t,\triangle f,\delta_T,\delta_R)+\\
%	&\sqrt{\frac{1}
%		{K_{RF}(t)+1}\cdot{\frac{1}
%			{K_{RF}(t+\triangle t)+1}}}R_{qp,p'q'}^{N}(t,f;\triangle t,\triangle f,\delta_T,\delta_R)	\end{aligned}
%	\end{equation}
\begin{figure*}[b]
	{\noindent} 
	\rule[-10pt]{18.07cm}{0.1em}	
	
	\begin{equation}
		\begin{aligned}
			R_{qp,q'p'}&(t,f;\triangle t,\triangle f,\delta_T,\delta_R)=E\left[H_{qp}(t,f)H_{q'p'}^*(t+\triangle t,f+\triangle f)\right]=
			\sqrt{\frac{K_{RF}(t)}
				{K_{RF}(t)+1}\cdot{\frac{K_{RF}(t+\triangle t)}
					{K_{RF}(t+\triangle t)+1}}}\\
			& \cdot R_{qp,q'p'}^{L}(t,f;\triangle t,\triangle f,\delta_T,\delta_R)+
			\sqrt{\frac{1}
				{K_{RF}(t)+1}\cdot{\frac{1}
					{K_{RF}(t+\triangle t)+1}}}R_{qp,q'p'}^{N}(t,f;\triangle t,\triangle f,\delta_T,\delta_R)	\end{aligned}
	\end{equation}
\end{figure*} 

\subsection{The Space CCF}
In terms of (3), we can get $h_{qp}(t)$ and $h_{q'p'}^*(t)$ easily. The space CCF can be denoted as
\begin{equation}
	\begin{split}
		\begin{aligned}
			&\rho_{qp,q'p'}(t;\delta_T,\delta_R)=E\left[h_{qp}(t)h_{q'p'}^*(t)\right]={\frac{K_{RF}(t)}{K_{RF}(t)+1}}\\
			&\cdot \rho_{qp,q'p'}^L(t; \delta_T,\delta_R)
			+{\frac{1}{K_{RF}(t)+1}}\cdot \rho_{qp,q'p'}^N(t; \delta_T,\delta_R).	\end{aligned}
	\end{split}
\end{equation}
% Besides, we can get space CCF by letting $\triangle t=0$ and $\triangle f=0$ in STFCF. 
\subsection{The Temporal ACF}
According to (3), we can get $h_{qp}(t)$ and $h_{qp}^*(t+\triangle t)$. The temporal ACF can be denoted as
\begin{equation}
	\begin{split}
		\begin{aligned}
			&r_{qp,qp}(t;\triangle t)=E\left[h_{qp}(t)h_{qp}^*(t+\triangle t)\right]=\\
			&\sqrt{\frac{K_{RF}(t)}
				{K_{RF}(t)+1}\cdot{\frac{K_{RF}(t+\triangle t)}
					{K_{RF}(t+\triangle t)+1}}}r_{qp,qp}^L(t; \triangle t)+\\
			&\sqrt{\frac{1}
				{K_{RF}(t)+1}\cdot\frac{1}
				{K_{RF}(t+\triangle t)+1}}r_{qp,qp}^N(t; \triangle t).	\end{aligned}
	\end{split}
\end{equation}
%Furthermore, the temporal ACF can be acquired by letting $q=q', p=p'$, and $\triangle f=0$ in STFCF. 
\subsection{The FCF}
In terms of (6), we can get $H_{qp}(t,f)$ and $H_{qp}^*(t,f+\triangle f)$. The FCF can be denoted as
\begin{equation}
	\begin{split}
		\begin{aligned}
			&\kappa_{qp,qp}(t,f;\triangle f)=E\left[H_{qp}(t,f)H_{qp}^*(t,f+\triangle f)\right]=\\
			&{\frac{K_{RF}(t)}
				{K_{RF}(t)+1}}\kappa_{qp,qp}^L(t,f;\triangle f)+{\frac{1}
				{K_{RF}(t)+1}}\kappa_{qp,qp}^N(t,f;\triangle f).	\end{aligned}
	\end{split}
\end{equation}
%\indent
%By setting $q=q', p=p'$, and $\triangle t=0$, the STFCF becomes the FCF.
\subsection{The SVS}
The channel matrix can be represented as singular value decomposition
\begin{equation}\label{capacity}
	\textbf{H}=\textbf{U}
	\textbf{$\Sigma$}
	\textbf{V}
\end{equation} where $\textbf{U}$ and $\textbf{V}$ are used to represent unitary matrixes, $\textbf{$\Sigma$}$ is used to represent K $\times$ M diagonal matrix. K and M denote the number of users and Tx antenna elements, respectively. Furthermore, the SVS can be calculated as
\begin{equation}\label{svd}
	\begin{aligned}
		\kappa_{\text{svs}}=\frac{\underset{k}{\max}\,    {\sigma_k}}{\underset{k}{\min}\,
			{\sigma_k}}
	\end{aligned}
\end{equation} where $\sigma_k$ ($k$=1, 2, $\cdots$, K) are the singular values, and $\kappa_{\text{svs}}$ is SVS.
\subsection{Channel Capacity}
Channel capacity is the maximum rate in channel where the bit error rate tends to zero. There are $M_T$ and $M_R$ antennas at Tx and Rx sides, respectively, and the Tx does not know the channel state information. If we choose signal covariance matrix as identity matrix $I_{M_T}$, which means the signals are independent and equi-powered at the transmit antennas, channel capacity can be represented as \cite{RE18}
\begin{equation}\label{capacity}
	C=\log_2\left[\det(I_{M_R}+\frac{\rho}
	{M_T}\textbf{H}\textbf{H}^H)\right]
\end{equation} where $\det\left[\cdot\right]$ defines the determinant, $(\cdot)^H$ defines the conjugate transpose operation, $I_{M_R}$ defines the identity matrix of size $M_R$, and $\rho$ defines the signal-to-noise ratio (SNR).

\section{Results and Analysis}
\begin{figure}[tb]
	\centerline{\includegraphics[width=0.4\textwidth]{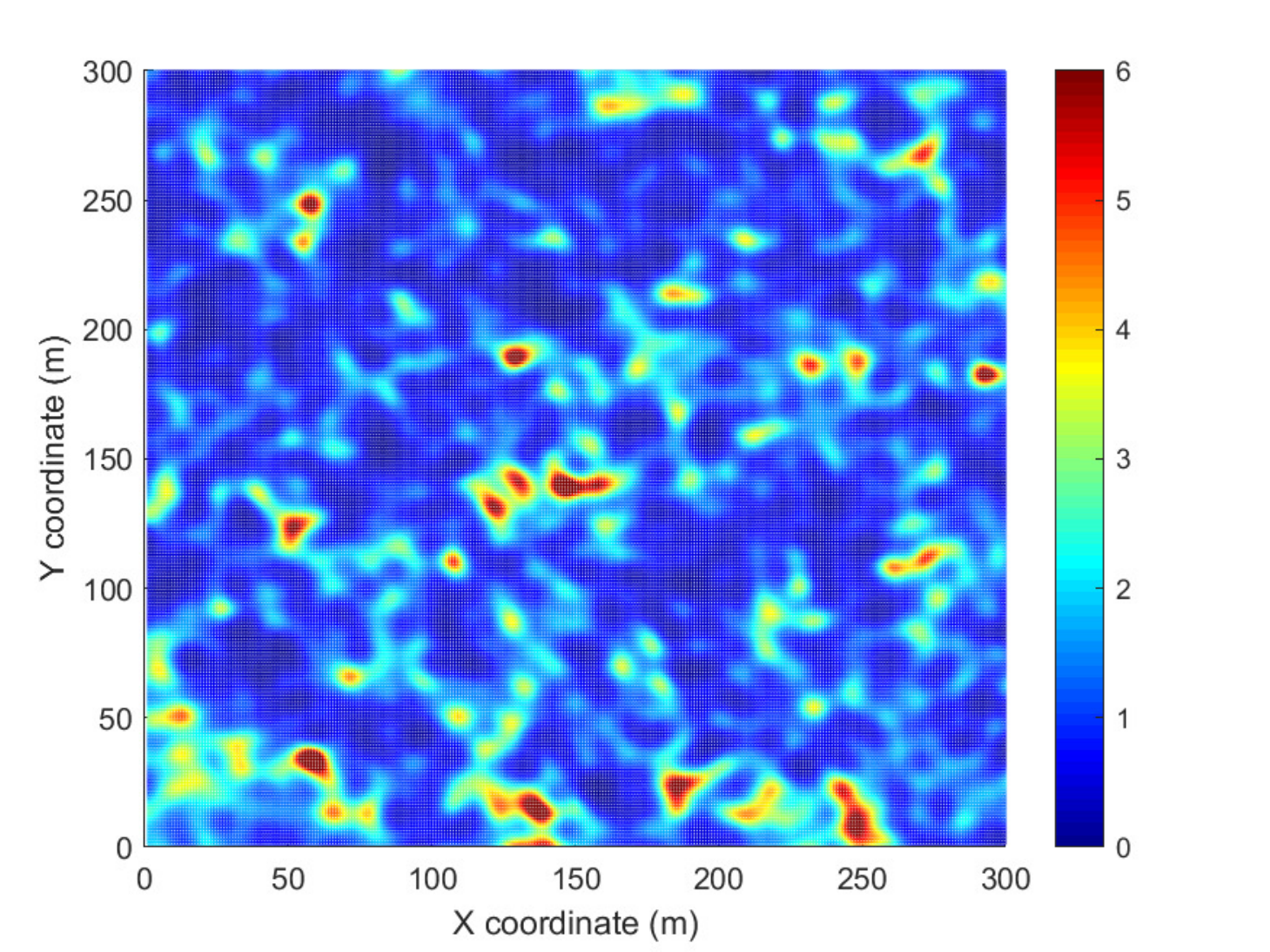}}		\caption{ Delay spread with spatial consistency in a 2D area.}
	\label{fig_2}
\end{figure}

The statistical properties and performance indicators of the model are simulated and analyzed in this section.
% It should be noted that the LSPs and SSPs are generated according to the positions of Tx and Rx, which makes the model inherently spatially consistent. 
LSPs with spatial consistency are generated through the SoS method. 
%	\cite{RE19}. 
As shown in Fig. \ref{fig_2}, it is the LSP of delay spread in 
an area of 300 $\text{m}$ $\times$ 300 $\text{m}$ and its parameters are set to 300 sine waves with the ACF modeled as a compound function of Gaussian and exponential decay. It can be seen obviously that the continuous spatial variation of delay spread factor is realized.

\subsection{The Temporal ACF}
\begin{figure}[b]
	\centerline{\includegraphics[width=0.4\textwidth]{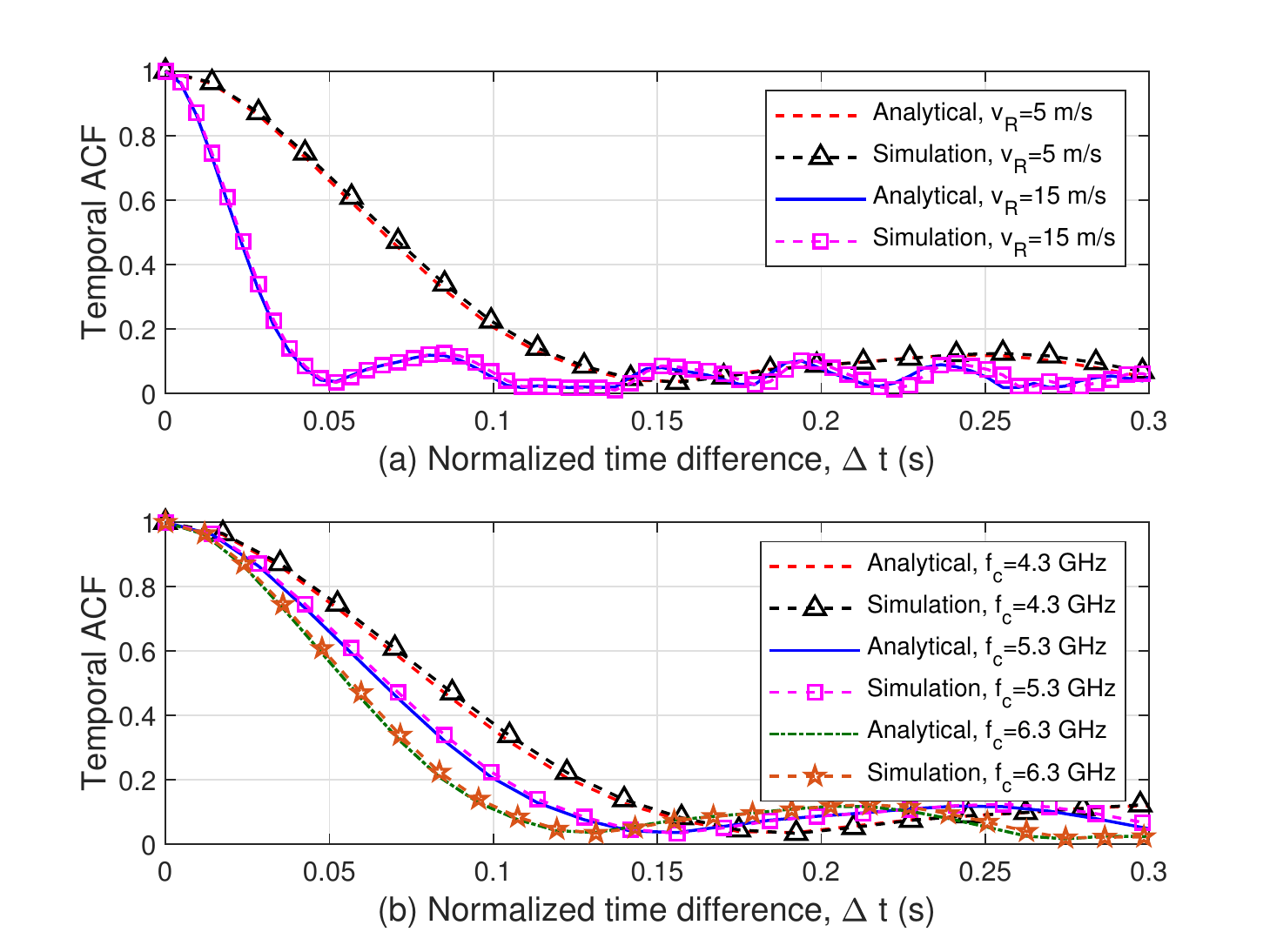}}		\caption{ (a) Temporal ACF with different velocities at Rx side (b) Temporal ACF with different carrier frequencies ($\beta_A^T$=$\pi$/10, $\beta_A^R$=$\pi$/12, $\alpha_A^T$=$\pi$/10, $\alpha_A^R$=$\pi$/12, $\lambda_G$=20/$\text{m}$, $\lambda_R$=1/$\text{m}$, $D_c^A$=40 $\text{m}$).}
	\label{fig_3}
\end{figure}

Fig. \ref{fig_3} illustrates the temporal ACF. Fig. \ref{fig_3} (a) represents the ACF changing with different velocities at Rx side. When the velocity at Rx side becomes larger, the coherence time will become shorter. The coherence time refers to the time difference when the ACF equals to a given threshold, which can be determined by system requirements. Fig. \ref{fig_3} (b) represents the ACF changing with different carrier frequencies. When the carrier frequency becomes larger, the coherence time will become shorter. The reasons for above phenomenon is that the larger velocity and carrier frequency lead to larger Doppler shift. Larger Doppler shift makes the channel more fluctuant and uncorrelated.
\subsection{The Space CCF}
\begin{figure}[tb]
	\centerline{\includegraphics[width=0.4\textwidth]{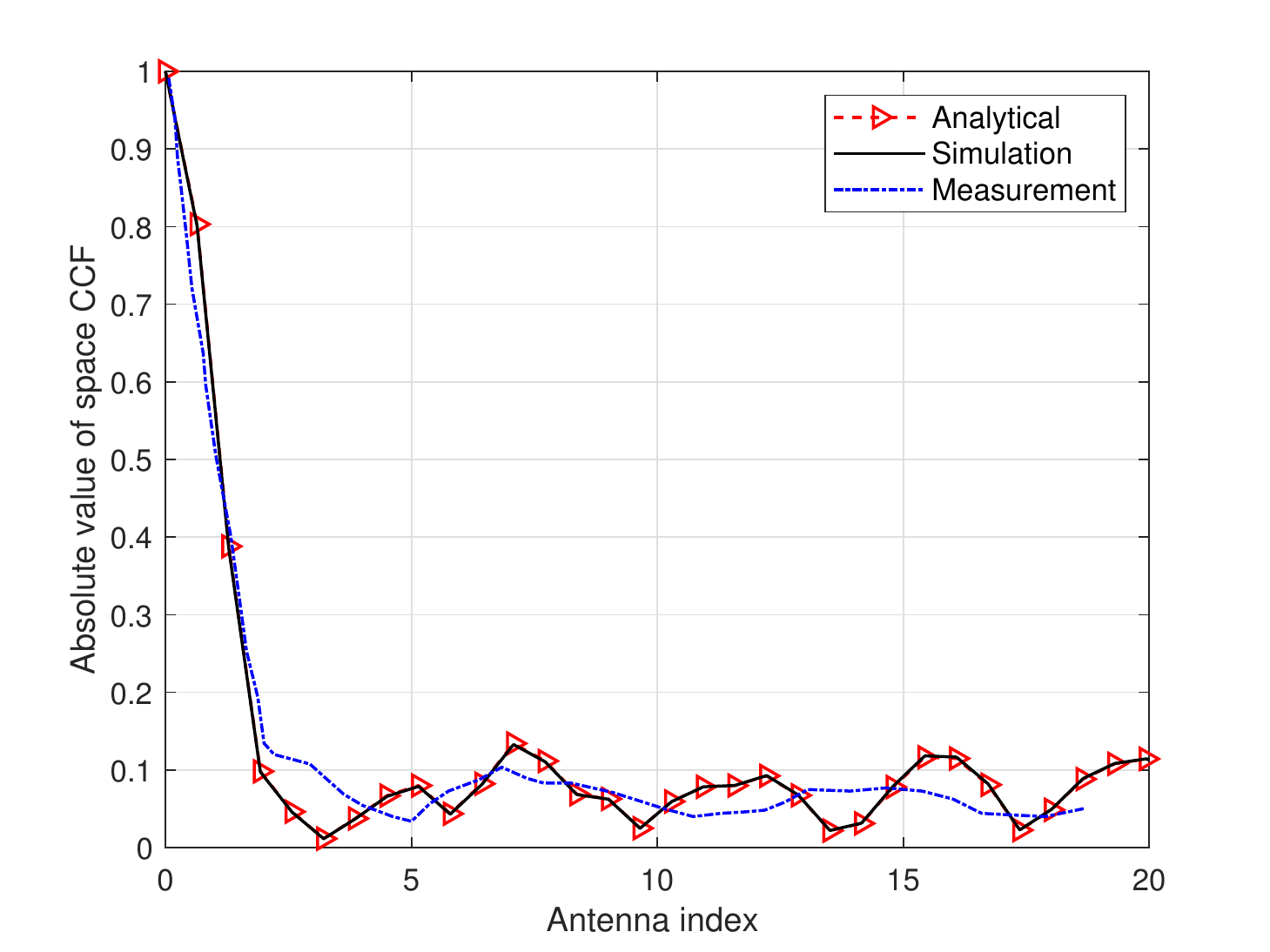}}		\caption{Absolute value of space CCF at different antenna spacings at the BS side ($f_c$=2.6 GHz, $M_T$=128, $M_R$=1, $\beta_A^T$=$\pi$/10, $\beta_A^R$=$\pi$/12, $\alpha_A^T$=$\pi$/10, $\alpha_A^R$=$\pi$/12, $\delta_T$=$\lambda/2$, $\lambda_G$=20/$\text{m}$, $\lambda_R$=1/$\text{m}$, $D_c^S$=40 $\text{m}$, $\text{NLOS}$). }
	\label{fig_4}
\end{figure}

Fig. \ref{fig_4} illustrates the space CCF. The measurement was conducted in a campus environment at 2.6 GHz carrier frequency with 128 antenna elements ULA at BS side \cite{RE20}. The simulation result is consistent with the measurement, which validated the presented model. 

\subsection{The FCF} 
\begin{figure}[tb]
	\centerline{\includegraphics[width=0.4\textwidth]{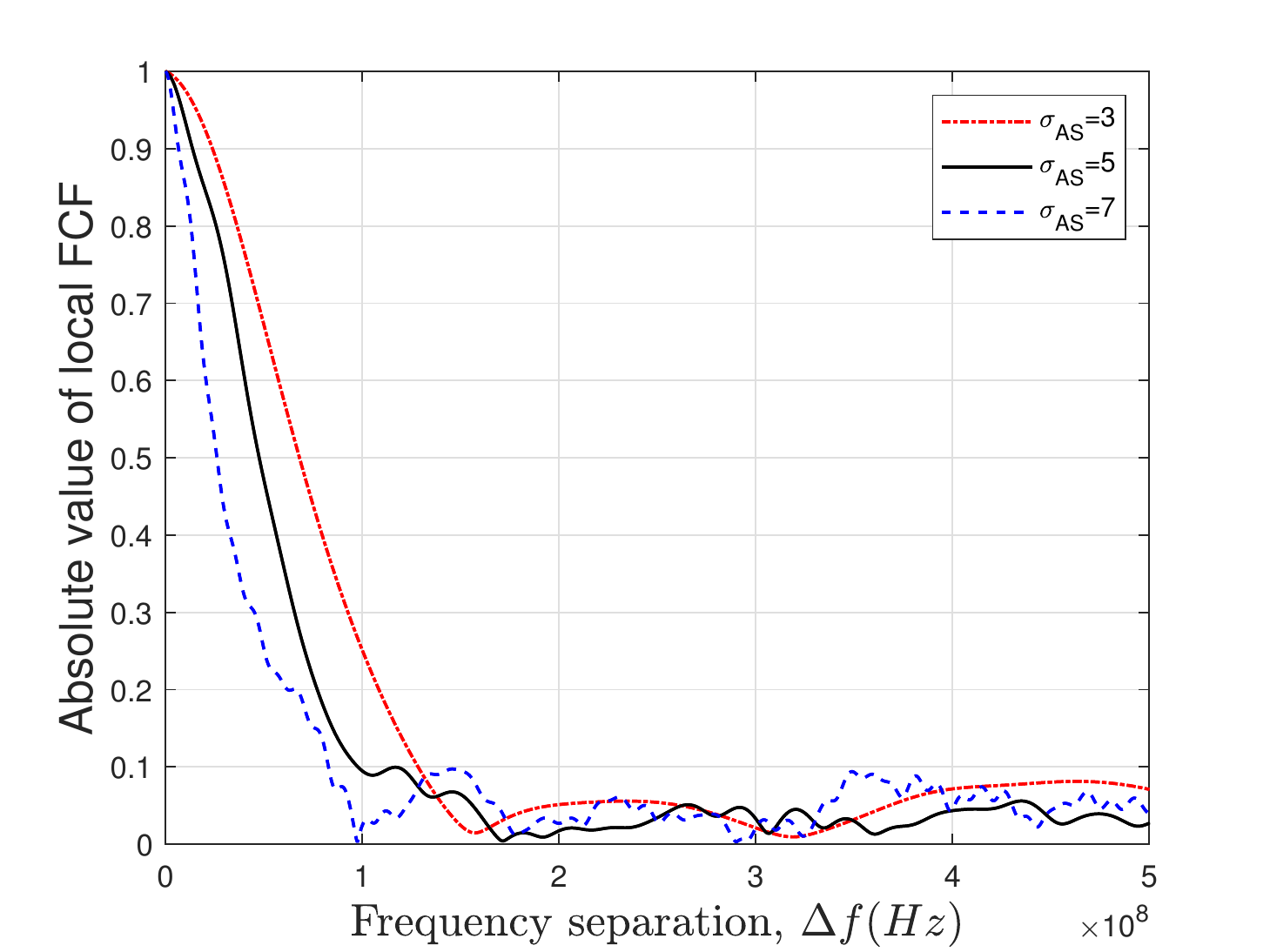}}		\caption{Absolute value of FCF with different cluster azimuth spreads ($f_c$=28 GHz, $\beta_A^T$=$\pi$/10, $\beta_A^R$=$\pi$/12, $\alpha_A^T$=$\pi$/10, $\alpha_A^R$=$\pi$/12, $\delta_T$=$\lambda/2$, $\lambda_G$=20/$\text{m}$, $\lambda_R$=1/$\text{m}$).}
	\label{fig_5}
\end{figure}

The FCF is shown in Fig. \ref{fig_5}. The channel with different cluster azimuth spread values 3, 5, and 7 has different coherence bandwidths 680 MHz, 475 MHz, and 260 MHz, respectively. What should be noted that is the coherence bandwidth refers to the frequency separation when the FCF equals to 0.5. The above phenomenon indicates that the larger cluster azimuth spreads will reduce the correlation of the channel. 
\subsection{STF Cluster Evolution} 
\begin{figure}[htbp]
	\centering{\includegraphics[width=0.45\textwidth]{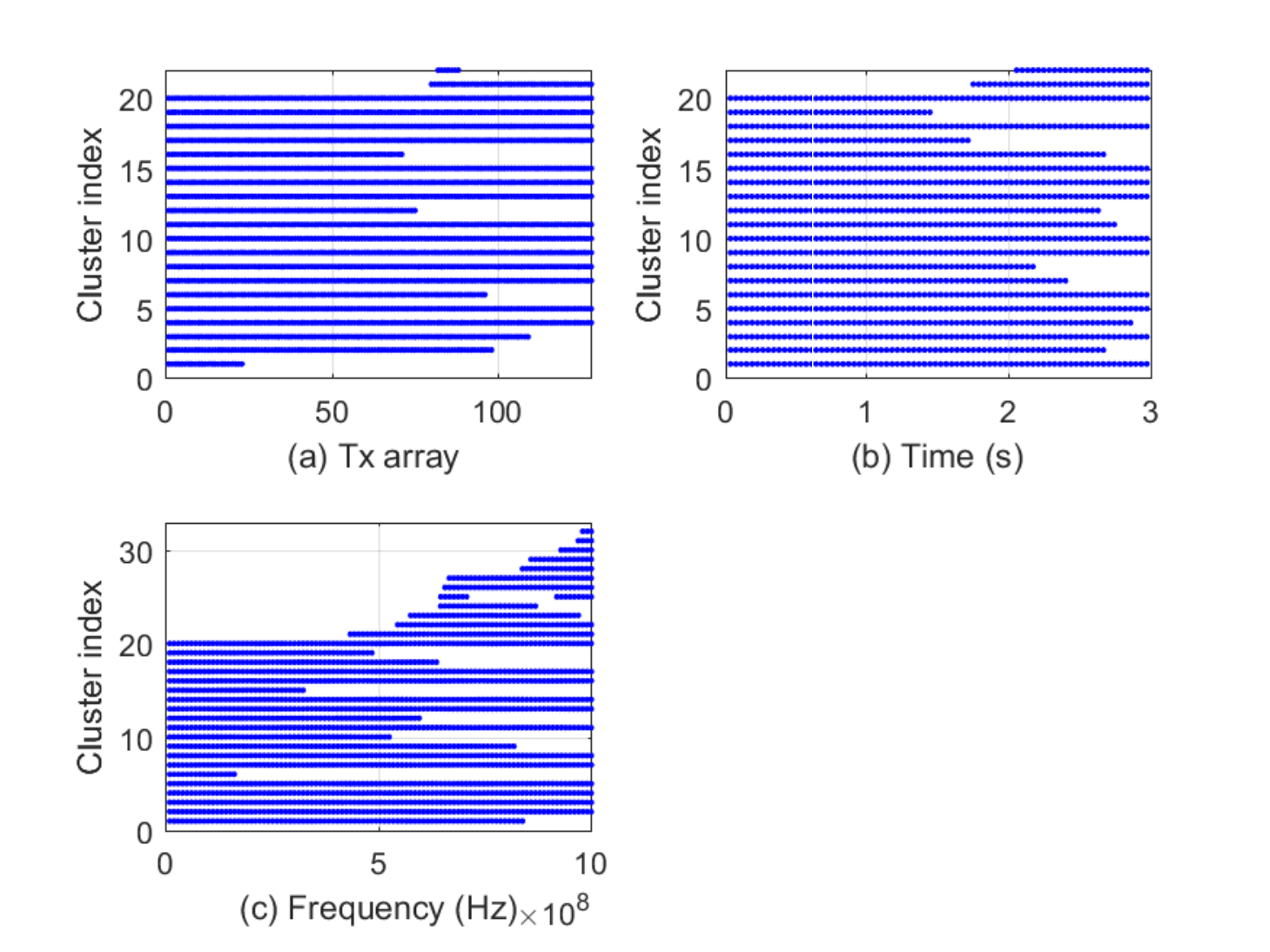}}
	\caption{(a) Cluster evolution in space domain ($f_c$=5.3 GHz, $M_T$=128, $\delta_T$=$\lambda/2$, $D_c^A$=40 m) (b) Cluster evolution in time domain ($f_c$=5.3 GHz, $v_R$=15 m/s, $v_T$=0 m/s, $D_c^S$=40 m) (c) Cluster evolution in frequency domain ($f_c$=38 GHz, $v_R$=15 m/s, $v_T$=0 m/s).}
	\label{fig_6}
\end{figure}

Cluster evolutions in STF domain are shown in Fig. \ref{fig_6} (a), Fig. \ref{fig_6} (b), and Fig. \ref{fig_6} (c). Different antennas at the same time instant and frequency, or the same antenna at different time instants and frequencies will see different clusters, which indicates the channel is non-stationary in STF domain. 
%    	\begin{figure}[tb]
%    		\centerline{\includegraphics[width=0.33\textwidth]{ArrEvo.pdf}}		\caption{Cluster evolution in space domain ($f_c$=5.3 GHz, $M_T$=128, $\delta_T$=$\lambda/2$, $D_c^A$=40 $\text{m}$ ).}
%    		\label{fig_6}
%    	\end{figure}
%    	\begin{figure}[tb]
%    		\centerline{\includegraphics[width=0.33\textwidth]{TimEvo.pdf}}		\caption{Cluster evolution in time domain ($f_c$=5.3 GHz, $v_R$=15 m/s, $v_T$=0 m/s, $D_c^S$=40 $\text{m}$).}
%    		\label{fig_7}
%    	\end{figure}
%    	\begin{figure}[tb]
%    		\centerline{\includegraphics[width=0.33\textwidth]{FreEvo.pdf}}		\caption{Cluster evolution in frequency domain ($f_c$=38 GHz, $v_R$=15 m/s, $v_T$=0 m/s).}
%    		\label{fig_8}
%    	\end{figure}    	
\subsection{The SVS}
\begin{figure}[tb]	\centerline{\includegraphics[width=0.4\textwidth]{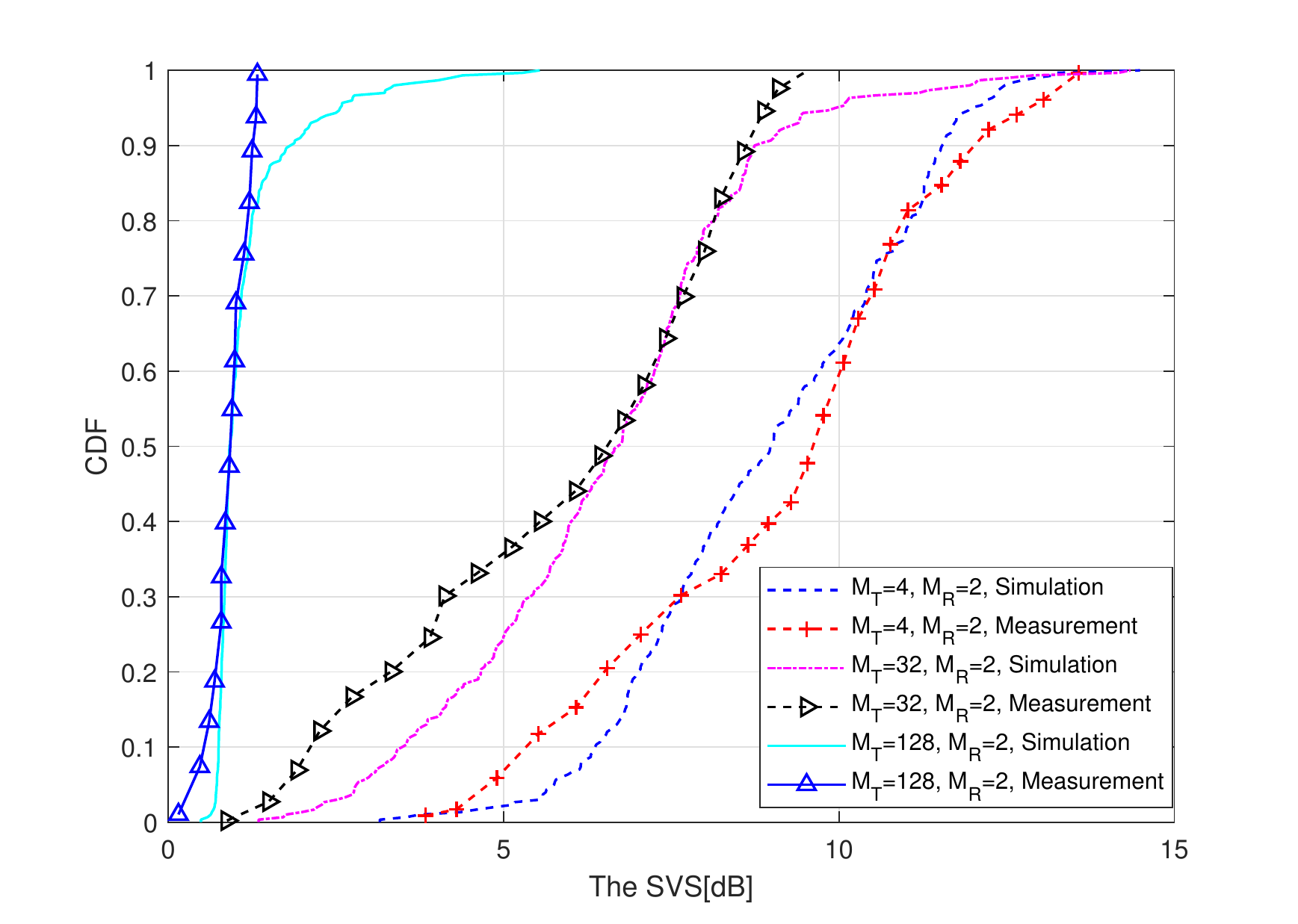}}		\caption{CDFs of SVSs with different numbers of antennas at Tx side ($f_c$=1.4725 GHz, $\beta_A^T$=$\pi$/3, $\beta_A^R$=$\pi$/4, $\alpha_A^T$=$\pi$/3, $\alpha_A^R$=$\pi$/3, $\delta_T$=$\lambda/2$, $\lambda_G$=20/$\text{m}$, $\lambda_R$=1/$\text{m}$, $\text{LOS}$).}
	\label{fig_7}
\end{figure}

Fig. \ref{fig_7} illustrates the cumulative distribution functions (CDFs) of SVSs of simulation results and measurement in \cite{RE21}. The channel measurement was performed in indoor scenario at 1.4725 GHz with a  virtual 128-element ULA. The simulation results agree with the measurement data. When $M_T$ gradually increases to 128, the SVS gradually decreases to below 1 dB. The above phenomenon manifests that the channel becomes more and more stable, and the channel vectors among users become approximately orthogonal. 
\subsection{Channel Capacity}
\begin{figure}[tb]
	\centerline{\includegraphics[width=0.4\textwidth]{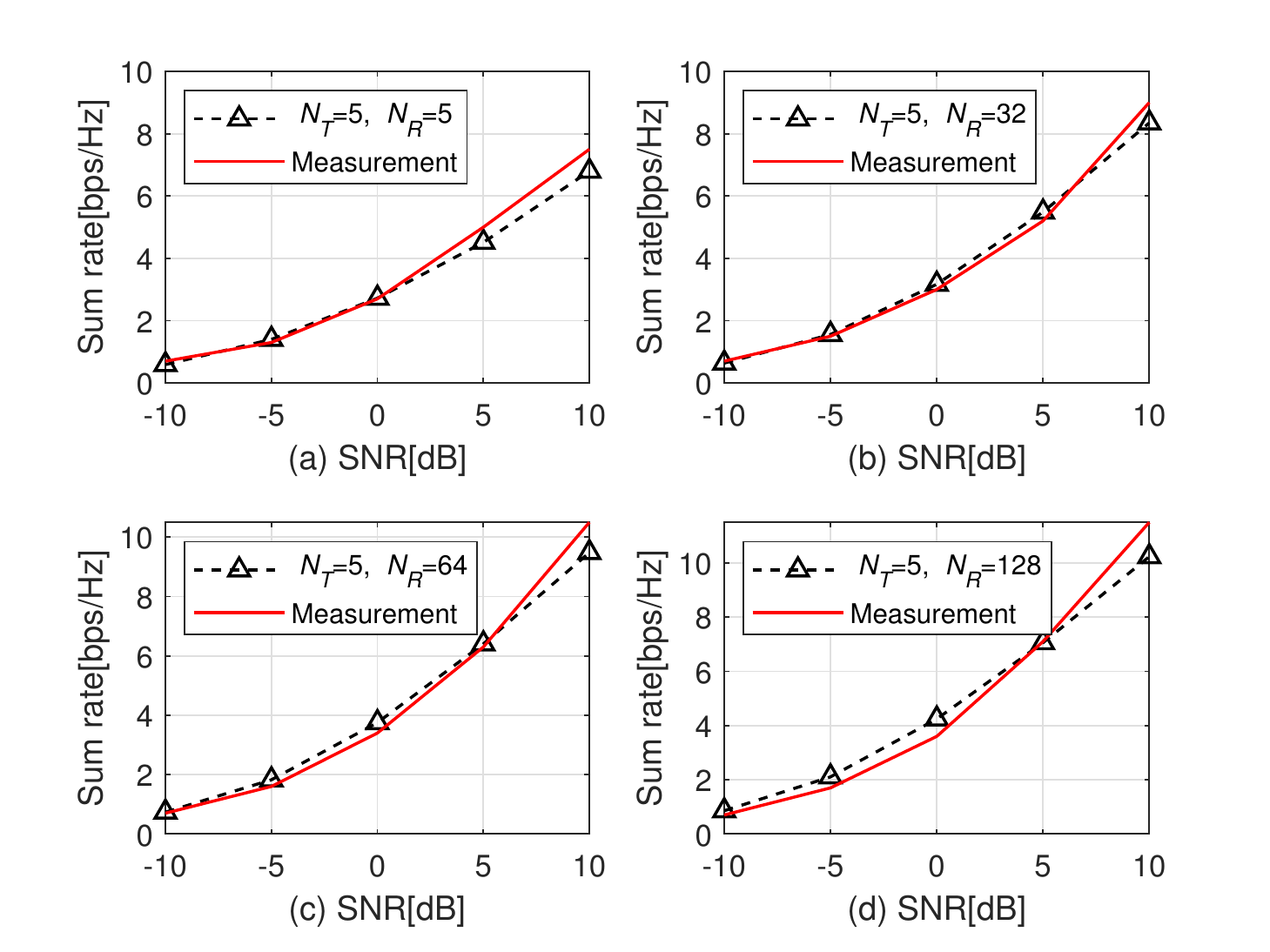}}		\caption{The uplink sum-rates with different numbers of antennas at Rx side ($f_c$=2.6 GHz,  $\beta_A^T$=$\pi$/10, $\beta_A^R$=$\pi$/12, $\alpha_A^T$=$\pi$/10, $\alpha_A^R$=$\pi$/12, $\delta_T$=$\lambda/2$, $\lambda_G$=20/$\text{m}$, $\lambda_R$=1/$\text{m}$).}
	\label{fig_8}
\end{figure}

The uplink sum-rates in the measured channels and the presented model are compared in Fig. \ref{fig_8} \cite{RE22}. The simulation results are consistent with the measurement data. With the number of antennas increasing at BS side, the uplink sum-rates also increase within a certain SNR range.
\section{Conclusions}
The paper has proposed a general 3D massive MIMO GBSM for 6G communication systems. The presented model can support arbitrary velocities and trajectories at both Tx side and Rx side, which are equipped with URAs. Meanwhile, it has studied cluster evolution in STF domain to support STF non-stationary communication scenario. In addition, the spatial consistency of LSPs generation has been proved by using the method of parameter generation with spatial consistency. The simulations about temporal ACF, space CCF, FCF, SVS, and channel capacity are conducted. Analytical, simulation results, and measurements have also been compared to verify the validity of the channel model. The novel GBSM proposed in this paper will play a significant part in the development of 6G communication systems.
\section*{Acknowledgment}
\small {This work was supported by the National Key R\&D Program of China under Grant 2018YFB1801101, the National Natural Science Foundation of China (NSFC) under Grant 61960206006 and Grant 61901109, the Frontiers Science Center for Mobile Information Communication and Security, the High Level Innovation and Entrepreneurial Research Team Program in Jiangsu, the High Level Innovation and Entrepreneurial Talent Introduction Program in Jiangsu, the Research Fund of National Mobile Communications Research Laboratory, Southeast University, under Grant 2020B01, the Fundamental Research Funds for the Central Universities under Grant 2242019R30001, the Huawei Cooperation Project, and the EU H2020 RISE TESTBED2 project under Grant 872172, the National Postdoctoral Program for Innovative Talents under Grant BX20180062.}

\end{document}